\documentclass[namedreferences]{solarphysics}
\usepackage[optionalrh]{spr-sola-addons}
\usepackage{epsfig}
\usepackage{graphicx}
\usepackage{color}
\usepackage{url}
\usepackage{lscape}
\usepackage{sidecap}
\newcommand{\etal}{{\it et al.}}
\usepackage[pdfborder={0 0 0 },urlcolor=blue]{hyperref}
\ifx \doiurl \undefined \def \doiurl#1{\href{http://dx.doi.org/#1}{\url{#1}}}\fi
\ifx \adsurl \undefined \def \adsurl#1{\href{http://adsabs.harvard.edu/abs/#1}{\url{#1}}}\fi

\begin{document}
\begin{article}
\begin{opening}

\title{North--South Asymmetry in Solar Activity and  Solar Cycle Prediction, 
IV: Prediction for Lengths of Upcoming Solar Cycles}

\author{J. Javaraiah}

\institute{\#58, 5th Cross, Bikasipura (BDA), Bengaluru-560 111,  India.\\
Formerly with Indian Institute of Astrophysics, Bengaluru-560 034, India.\\
email: \url{jajj55@yahoo.co.in;  jdotjavaraiah@gmail.com}\\
}

\runningauthor{J. Javaraiah}
\runningtitle{Prediction for Lengths of Upcoming Solar Cycles}

\begin{abstract}
 We analyzed the daily sunspot-group  data  reported by the  Greenwich
 Photoheliographic Results (GPR) during the period 1874\,--\,1976 and
 Debrecen Photoheligraphic Data (DPD) during the period 1977\,--\,2017
 and studied North--South asymmetry in the maxima
and minima  of the Solar Cycles~12\,--\,24.
We derived the time-series of the 13-month smoothed monthly
mean corrected whole-spot areas of the sunspot groups in the Sun's whole sphere
(WSGA), northern hemisphere (NSGA), and southern hemisphere (SSGA). From these
smoothed time series we obtained the values of the maxima and minima, and the
corresponding epochs, of the WSGA, NSGA, and SSGA Cycles 12\,--\,24.
 We find that there exists  a 44\,--\,66 years periodicity in the 
North--South  asymmetry of minimum. A long periodicity (130\,--\,140 years)
 may exist in   the asymmetry of maximum.
 A statistically significant correlation exists between
  the maximum of  SSGA Cycle~$n$  and the rise time of  WSGA Cycle~$n+2$.
A   reasonably significant correlation also exists  between
 the maximum of  WSGA Cycle~$n$  and the decline time of 
 WSGA Cycle~$n+2$. These relations   suggest
that the solar dynamo carries memory over at least three solar cycles.
Using these relations we obtained 
  the values $11.7\pm 0.15$ years, $11.2\pm 0.2$ years,
and $11.45\pm 0.3$ years for the lengths of WSGA Cycles 24, 25, and 26, 
respectively, and hence,  July 2020, October 2031,  and March 2043 for 
the minimum epochs (start dates) of   WSGA Cycles 25, 26, and 27, respectively.
We  also obtained   May 2025 and  March 2036     
for the maximum epochs of  WSGA Cycles 25 and 26, respectively.
 It seems during the late Maunder Minimum sunspot 
 activity was absent around the epochs of the maxima of  NSGA-cycles
and the minima of  SSGA-cycles, and some activity was present at the 
epochs of the maxima of some SSGA-cycles and the minima of some NSGA-cycles.
\end{abstract}


\end{opening}

\section{Introduction}
North--South asymmetry in solar activity may  have
 some important implications for the solar-dynamo
 mechanism ($e.g.$ \opencite{soko94}; \opencite{bd13}; \opencite{shetye15}).
Therefore, a  study of variations in the North--South asymmetry of solar
 activity
 may be useful for better understanding the underlying mechanism of
the solar cycle. Besides a 11\,--\,12-year
 periodicity (\opencite{carb93}; \opencite{jg97a}) and many short
 periodicities (\opencite{vizo90}; \opencite{chow13}; \opencite{rj15}, 
and references therein), the existence
of a $\approx$110-year periodicity~(\opencite{verma93}) in the 
North-South asymmetry of solar activity is known. It is well known that
 large and small sunspot groups (in general active regions) have
 different physical 
origin~(\opencite{pf72}; \opencite{gf79}; \opencite{howard96};
 \opencite{jg97b}; \opencite{kmh02}; \opencite{siva03}). 
  Recently, short-term periodicities in
the variations of the North--South asymmetry in the numbers of different-size 
sunspot groups have also been studied~(\opencite{mb16}). \inlinecite{verma93}
 has found the $\approx$110-year periodicity in the North--South asymmetry of
solar activity by using the combined data of many different indices  of
solar activity that were available  for
different periods: sunspot area (1832\,--\,1976), sunspot counts 
(1833\,--\,1877), sunspot groups (1954\,--\,1986),
 solar flares (1936\,--\,1954), major flares (1955\,--\,1979),
 gamma-ray flares (1980\,--\,1986), and H$\alpha$ flares (1987\,--\,1990).
Here we study the long-term variation in the North--South
asymmetry of solar activity by using the 143 years of updated  GPR
and DPD  sunspot group data, $i.e.$ by using the  data
of a single activity index. The average value of the areas of
 sunspot groups in the solar maximum  is much larger than that of the sunspot
 groups in the solar minimum.  In addition, it seems the  maxima and minima
 of solar cycles comprise relatively strong quadrupole and dipole magnetic 
fields, respectively~($e.g.$ \opencite{dbh12}). 
  Hence, the cycle-to-cycle modulations in 
maximum and minimum could be different and  may have different implications
 on the solar dynamo.
 Therefore, here we have examined the  long-term variations
in the North--South asymmetry of solar maximum and minimum,
independently.

Prediction of the properties of  solar cycles is important for understanding 
the underlying mechanism of the solar cycle. Prediction of the amplitude and 
the length (period) of a solar cycle is  important also for monitoring 
 space weather  and for understanding the solar--terrestrial relationship. 
A wide range of techniques have been used for prediction of the amplitude of 
a solar cycle~(\opencite{kane07}; \opencite{pes08}; \opencite{obrid08}; 
\opencite{hath15}).
Although the average cycle length can be fairly accurately
 determined~(\opencite{hath15}), so far no technique is available 
 to make  prediction of the maximum epoch and  length  of a cycle.  
By using the well-known  Waldmeier 
effect~(\opencite{wald35}; \opencite{wald39}; \opencite{hath15}), 
$i.e.$ inverse relationship between  the amplitude of a cycle and 
its rise time (the time taken for  rising from minimum to
 maximum),  it is only  possible to make 
an approximate prediction for the rise time of a cycle from a predicted
 amplitude.
The existence of an approximate inverse relationship between the 
amplitude of a cycle and its length (time from the minimum of a cycle 
to the minimum of next cycle) is also well-known~(\opencite{hs97}; 
\opencite{solanki02}). By using this relationship it is only possible 
to make  an approximate prediction for the length of a cycle from the 
known/predicted amplitude. 

 The solar rotational and meridional flows  may cause the magnetic 
fields at different heliographic latitudes during
different time intervals of  a solar cycle to contribute to the
 activity at the same or different heliographic latitudes during the
following cycle(s). Hence, in  our earlier articles  
(Article~I : \opencite{jj07}, Article~II: \opencite{jj08}, 
and Article~III: \opencite{jj15}) 
the combined Greenwich and {\it Solar Optical Observing Network} (SOON)
 sunspot-group data during 1874\,--\,2013 were analyzed and
  correlations 
of  different latitude bands of activity (sums of the areas of sunspot 
groups) with the amplitude of solar cycle were determined. 
 It was found that the sums of 
the areas of the sunspot groups in the $0^\circ$\,--\,$10^\circ$ 
latitude interval
 of the Sun's northern hemisphere and in the time interval of $-1.35$ year to
 $+2.15$ year from the time of the preceding minimum of a sunspot cycle,  
and also in the same latitude interval of the southern hemisphere but
 $+1.0$ year to $+1.75$ year from 
the time of the maximum of a sunspot cycle, correlate well with the amplitude 
(maximum of the 13-month smoothed monthly sunspot number) of its immediate 
following cycle. 
 This enabled us to made predictions for the amplitude of Solar Cycle~24 
(\opencite{jj07}, \citeyear{jj08}),  and
 that of Cycle~25 (\opencite{jj15}). 
   Our predictions ($74\pm10$ and $87\pm7$) for the amplitude 
of Cycle~24   closely agree with   the observed amplitude (81.9).   
 We have predicted $50\pm10$ for the amplitude of Solar Cycle~25 
(\opencite{jj15}). 
 
 A number of authors
predicted the amplitude and the maximum epoch of Cycle~25 by using
various techniques (\opencite{pk18}; \opencite{sarp18}, and references therein).
 Some authors predicted a high amplitude (up to $\approx$150) and some others  
 predicted a low ($\approx$50)
amplitude and years 2023\,--\,2025 for the maximum epoch of this cycle.
 Here we have  attempted to predict  
the maximum epochs of Solar Cycles 25 and 26 and the lengths of 
Cycles 24, 25, and 26. As mentioned  above,   
 in  our earlier articles 
 we have used the North--South asymmetry in the latitude distribution of 
sunspot activity. Here, we used the North--South asymmetry property of the 
maxima and minima of the solar cycles. 
 However, the  physical mechanism (flux-transport-processes) behind  
both our  earlier and  present  methods is same. 

In the next section we describe the data analysis, in Section~3 we present
the results,  and in Section~4 we summarize the conclusions and discuss
briefly.

\section{Data analysis}
The daily data of sunspot groups in the Greenwich
 Photoheliographic Results (GPR) during the period 
April 1874\,--\,December 1976 and
 Debrecen Photoheliographic data (DPD) during the period 
January 1977\,--\,June 2017 
are downloaded from
{\sf fenyi.solarobs.unideb.hu/pub/DPD/}
(for details see \opencite{gyr10}). 
These data contain the date with time,  heliographic latitude 
and longitude, corrected whole-spot area [A] in millionth of 
solar hemisphere (msh), central meridian 
distance, {\it etc.} of each of the sunspot groups observed on each day. 
The time series of the 13-month smoothed  monthly mean values 
of the international sunspot number  $R_{\rm Z}$ (ISSN) and that of  the   
 corresponding revised sunspot number (SN) 
during the period 1874\,--\,2017 are  downloaded  from 
{\sf www.sidc.be/silso/datafiles}
(Source: WDC-SILSO, Royal Observatory of Belgium, Brussels).
From the  13-month smoothed monthly mean   $R_{\rm Z}$-values 
 we obtained the maximum [$R_{\rm M}$]  and 
minimum [$R_{\rm m}$] values of the Sunspot (ISSN) Cycles 12\,--\,24 and 
their corresponding epochs. 
 We determined separate time series of 
13-month smoothed monthly mean corrected whole-spot areas
 of the  sunspot groups occurred in the Sun:  whole sphere (WSGA),  
northern hemisphere (NSGA),  and   southern hemisphere (SSGA), in 
 exactly the same way as the corresponding smoothed time
 series of $R_{\rm Z}$ are determined.  
 That is, first we determined the   
average value of the corrected areas of whole sunspot groups 
in each of the classes WSGA, NSGA, and SSGA  for  each calender
 month  of each year during the period 1874\,--\,2017, and then we    
  determined  the corresponding time series of 
 13-month smoothed monthly mean values.
 From these smoothed time series  we obtained the values of the 
 maxima and minima,  and their corresponding epochs, of the  WSGA, NSGA, and
 SSGA  Cycles 12\,--\,24.
We also determined the rise times of all the Cycles 12\,--\,24, and
the decline times and the lengths  
of the Cycles 12\,--\,23.
The North--South asymmetry in the maximum value and that in the minimum 
value of each cycle are calculated. 
The epochs of the  maxima  and 
minima of the cycles of $R_{\rm Z}$ and  the sunspot area are 
 not exactly the same   in the case of  some solar cycles 
($e.g.$ \opencite{ramesh08}). In fact,  
\inlinecite{dgt08} found that Waldmeier effect is not present  in the case of
 solar cycles in the area of sunspot groups. 
In an earlier analysis (\opencite{jj12})  we  found that for many
cycles the positions of the maxima of the small, large (medium-sized),
and very-large sunspot groups are different  and also considerably different
 from the corresponding ISSN maxima. Therefore, 
 from  the 13-month smoothed monthly mean time series 
  we have determined  the values of the  actual maximum [$A_{\rm M}$] and 
minimum [$A_{\rm m}$]  and their corresponding epochs, and also 
 the values [$A^*_{\rm M}$ and $A^*_{\rm m}$]  
 corresponding to the epochs of maximum [$R_{\rm M}$]  and minimum 
[$R_{\rm m}$] of the ISSN cycle. 
To make predictions for the rise times and the lengths of the upcoming 
solar cycles we study cross-correlations   among the parameters of 
 the WSGA, NSGA, and SSAG  cycles.
All of the  aforementioned  calculations
 were done by using 
the epochs of $R_{\rm M}$  and $R_{\rm m}$ of the sunspot cycles obtained from
 both the original (ISSN) and the revised (SN) time series. However, the    
 results  correspond to the epochs of $R_{\rm M}$  and $R_{\rm m}$ of 
 the revised  and the original time series  were 
found to be  almost the same. Hence, in the next section we report   
the results derived by using  the original  time series  only.

\begin{figure}
\centering
\includegraphics[width=\textwidth]{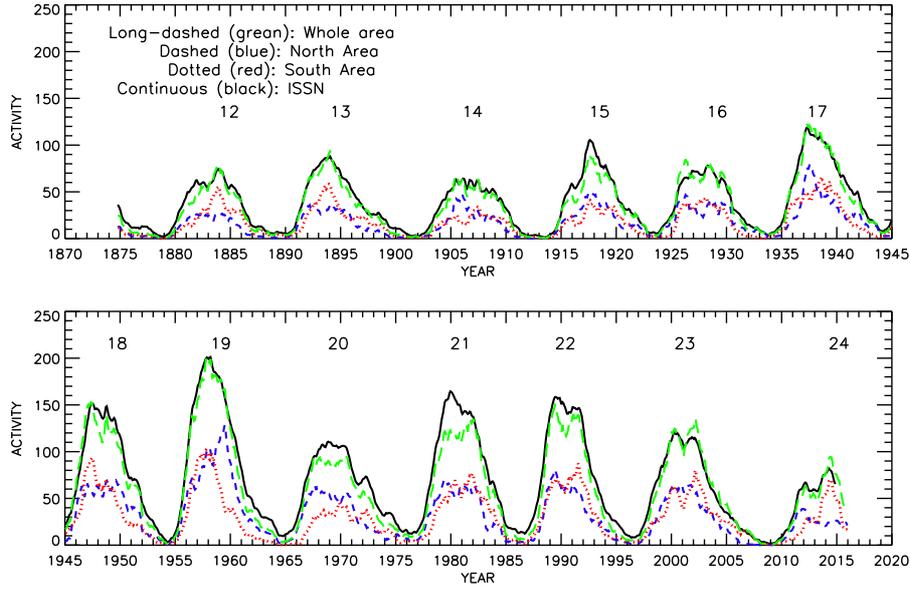}
\caption{Variations in the 13-month smoothed monthly mean area  of sunspot
 groups in the whole hemisphere: WSGA ({\it green-long-dashed curve}),
 northern hemisphere: NSGA ({\it blue-dashed curve}), southern hemisphere: SSGA 
({\it red-dotted curve}),  and international
sunspot number (ISSN) $R_{\rm_Z}$ ({\it black-continuous curve}).
 The values of WSGA, NSGA, and SSGA 
 are first divided by the largest value of WSGA, 3480.15 msh, and then 
multiplied by  
the largest value, 201.3, of ISSN. Waldmeier numbers of the solar cycles 
are also shown.}
\label{f1}
\end{figure}

\section {Results}
\subsection{Long-Term Variation in North--South Asymmetry}
Figure~1 shows the variations in the 13-month smoothed  monthly means
of WSGA, NSGA, and SSGA.   In this 
figure the corresponding variation in the $R_{\rm Z}$ is also shown. 
For  Solar Cycles 12\,--\,24 in Table~1 we have given the values 
 of $A^*_{\rm M}$ and $A^*_{\rm m}$ determined from the data of 
 WSGA, NSGA, and SSGA.
 In this table we have also given the 
values of $R_{\rm M}$ and $R_{\rm m}$   and their epochs.
 In Table~2 we have given the  values of the actual  maximum 
 [$A_{\rm M}$] and  minimum [$A_{\rm m}$],  and their epochs, of
WSGA, NSGA, and SSGA cycles. 
 In this table the values  of $A^*_{\rm S}$ ($A^*_{\rm N}$), $i.e.$
 the   13-months smoothed monthly mean SSGA (NSGA)
at the epochs  of the actual  maxima
and minima of NSGA (SSGA) cycles are also given.
In Table~3 we have given the rise time (RT), decline time (DT), and
 length (L) of ISSN, WSGA, NSGA, and SSGA cycles.
As can be seen in Figure~1, as already known, in  some solar cycles  
the positions of the maximum  peaks in the sunspot number and area data
 are reasonably different. The existence of double (or multiple)   
 peaks~(\opencite{gnevy63})  is clearly seen in most of the 
cycles of both NSGA and SSGA and  hence, the results are  consistent with 
the conclusion of
 \inlinecite{ng10} that the Gnevyshev gap (one to two year gap between
 the Gnevyshev peaks)  occurs
in both hemispheres and is not, in general, due to the superposition of 
two hemispheres out of phase with each other.    
In Table~4 we have given the difference between the epochs  of
 maxima and minima of the ISSN  and WSGA cycles, and between 
the epochs of maxima and minima of NSGA and SSGA cycles.
  Figures~2 and 3 show the cycle-to-cycle variations in the values 
 of the   maxima and  minima given in Tables~1 and 2. 
 As can be seen in Figures~2(a) and 3(a) and Table~4, 
 the epochs of $A_{\rm M}$ and  $A_{\rm m}$ are not exactly the same as the
corresponding epochs of   $R_{\rm M}$ and  $R_{\rm m}$. In Cycles~16  
 $A_{\rm M}$ led  $R_{\rm M}$ by 2.0 years, 
whereas in Cycles~20, 21, and 23 $R_{\rm M}$ lead
  $A_{\rm M}$ by 1.66, 1.75, and 1.91 years, respectively. In spite of the 
phase difference, there exists a good correlation (correlation coefficient
 $r = 0.95$) between  $A_{\rm M}$ and $R_{\rm M}$. However,  the 
$A_{\rm M}$ of Cycle~21 is found to be  smaller than 
that of Cycle~22 (a clear violation of the tentative inverse
 Gnevyshev--Ohl rule, see \opencite{jj17}), 
whereas $R_{\rm M}$ of Cycle~21 is slightly larger than that of Cycle~22, 
indicating that Cycle~21 consists of relatively large number of small sunspot
 groups than Cycle~22. 
This is also same  in the cases of the corresponding 
 NSGA and SSGA cycles (see Figure~2(b)). North-south asymmetry in solar activity
 may have a significant contribution to the violation of the
Gnevyshev--Ohl rule~(\opencite{jj16}).

\begin{SCtable}
{\scriptsize
\caption[]{The values [msh] of $A^*_{\rm M}$ and $A^*_{\rm m}$, $i.e.$   
 the   13-month
 smoothed monthly mean  areas  of the sunspot
 groups in whole sphere (WSGA),   
 northern hemisphere (NSGA), and 
 and southern hemisphere (SSGA),  
at the epochs $T_{\rm M}$ and $T_{\rm m}$ of the  maxima and minima  of 
the Sunspot (ISSN) Cycles 12\,--\,24. The values of the 
 maximum [$R_{\rm M}$] and 
 minimum [$R_{\rm m}$] of the ISSN cycles are also given.} 
\begin{tabular}{lccccccccc}
\noalign{\smallskip}
\hline
Cycle & \multicolumn{2}{c}{ISSN} & \multicolumn{1}{c}{WSGA} &
\multicolumn{1}{c}{NSGA}&\multicolumn{1}{c}{SSGA}\\
  \noalign{\smallskip}
\cline{2-3}
\noalign{\smallskip}
 &$T_{\rm M}$ & $R_{\rm M}$ &$A^*_{\rm M}$ &  $A^*_{\rm M}$&  $A^*_{\rm M}$\\ 
\noalign{\smallskip}
\hline
\noalign{\smallskip}
 12& 1883.958 &  74.6& 1370.73&  413.94&  956.80\\
 13& 1894.042 &  87.9& 1616.02&  621.13&  994.90\\
 14& 1906.123 &  64.2& 1043.95&  761.15&  282.80\\
 15& 1917.623 & 105.4& 1535.36&  828.54&  706.82\\
 16& 1928.290 &  78.1& 1323.98&  630.83&  693.15\\
 17& 1937.288 & 119.2& 2119.51& 1308.77&  810.73\\
 18& 1947.371 & 151.8& 2641.41& 1051.23& 1590.18\\
 19& 1958.204 & 201.3& 3441.37& 1748.74& 1692.63\\
 20& 1968.874 & 110.6& 1556.31&  951.02&  605.29\\
 21& 1979.958 & 164.5& 2121.16& 1063.77& 1057.40\\
 22& 1989.538 & 158.5& 2569.48& 1340.27& 1229.21\\
 23& 2000.290 & 120.8& 2152.54& 1110.94& 1041.60\\
 24& 2014.288 &  81.9& 1599.82&  419.47& 1180.35\\
\noalign{\smallskip}
\noalign{\smallskip}
 &$T_{\rm m}$ & $R_{\rm m}$ &$A^*_{\rm m}$ &  $A^*_{\rm m}$&  $A^*_{\rm m}$\\ 
\noalign{\smallskip}
 12& 1878.958&    2.2&   16.22&   10.55&    5.67\\
 13& 1890.123&    5.0&   68.99&   21.99&   47.00\\
 14& 1902.042&    2.7&   29.17&   16.66&   12.51\\
 15& 1913.455&    1.5&   13.22&    5.25&    7.98\\
 16& 1923.538&    5.6&   50.59&   30.22&   20.37\\
 17& 1933.707&    3.5&   32.31&   29.70&    2.61\\
 18& 1944.124&    7.7&  105.32&   65.53&   39.79\\
 19& 1954.288&    3.4&   24.14&    4.44&   19.71\\
 20& 1964.791&    9.6&   49.45&   45.35&    4.10\\
 21& 1976.206&   12.2&  149.70&   77.91&   71.78\\
 22& 1986.707&   12.3&   91.77&   62.41&   29.35\\
 23& 1996.373&    8.0&   88.07&   25.78&   62.29\\
 24& 2008.958&    1.7&    6.57&    4.57&    2.01\\
 \noalign{\smallskip}
\hline
\end{tabular}
\label{table1}
}
\end{SCtable}

As can be seen in Figure~2(b) in  most of the cycles  $A_{\rm M}$ of the  
northern hemisphere led that of the southern hemisphere by about two years. 
However, in the strong Cycles~18 and 19, $A_{\rm M}$ of the southern hemisphere 
led that of the northern by about two years (see Table~4). That is, 
the maximum is not always occurring first in northern hemisphere activity.  
As can be seen in Figure~3(b) in several cycles  $A_{\rm m}$ of the
northern hemisphere led that of the southern hemisphere by about one year,  
whereas in Cycles~21 and 23 $A_{\rm m}$ of the southern hemisphere 
led that of the northern hemisphere by about one year. 
The correlation between $A_{\rm m}$ (of WSGA) 
and $R_{\rm m}$  is also good ($r= 0.88$) but it is 
   less than that of  $A_{\rm M}$ 
and $R_{\rm M}$.

\begin{table}
{\scriptsize
\caption[]{The maximum   [$A_{\rm M}$] and
 the minimum [$A_{\rm m}$] 
 values [msh] and their corresponding 
 epochs  of the WSGA, NSGA, and SSGA Cycles 12\,--\,14, determined from the  
 corresponding time series of 13-month smoothed monthly
 mean areas of the sunspot groups in whole sphere,  
 northern hemisphere, and southern hemisphere, respectively.
 The values [msh] of $A^*_{\rm S}$ ($A^*_{\rm N}$), $i.e.$
 the   13-month smoothed monthly mean SSGA (NSGA)
at the epochs  of the   maxima
and minima of NSGA (SSGA) cycles are also given.}
\begin{tabular}{lccccccccccc}
\noalign{\smallskip}
\hline
Cycle & \multicolumn{2}{c}{WSGA} && \multicolumn{3}{c}{NSGA} &&
\multicolumn{3}{c}{SSGA}\\
  \noalign{\smallskip}
\cline{2-3}
\cline{5-7}
\cline{9-11}
\noalign{\smallskip}
 &Year & $A_{\rm M}$ &&Year & $A_{\rm M}$&$ A^*_{\rm S}$ && Year  & $A_{\rm M}$&$ A^*_{\rm N}$\\
\noalign{\smallskip}
\hline
\noalign{\smallskip}
 12& 1883.958& 1370.73&& 1882.371&  476.41&  501.19  && 1883.958& 956.80& 413.94\\
 13& 1894.042& 1616.02&& 1894.123&  649.18&  935.37  && 1893.958&1007.68& 577.00\\
 14& 1905.455& 1160.98&& 1906.042&  821.94&  287.14  && 1907.455& 585.47& 512.49\\
 15& 1917.538& 1554.25&& 1917.707&  854.59&  637.61  && 1917.623& 706.82& 828.54\\
 16& 1926.288& 1466.97&& 1926.288&  808.82&  658.16  && 1928.206& 754.08& 583.79\\
 17& 1937.288& 2119.51&& 1937.538& 1373.87&  714.09  && 1938.623&1140.50& 714.67\\
 18& 1947.371& 2641.41&& 1949.623& 1200.13&  860.11  && 1947.455&1614.13& 1011.14\\
 19& 1957.958& 3480.15&& 1959.538& 2220.33&  626.72  && 1957.790&1754.90& 1673.30\\
 20& 1970.538& 1627.50&& 1967.623& 1108.01&  488.26  && 1969.874& 782.85& 766.50\\
 21& 1981.707& 2338.33&& 1979.371& 1210.92&  855.77  && 1981.958&1317.10& 975.81\\
 22& 1989.455& 2591.13&& 1989.455& 1401.04& 1190.09  && 1991.538&1511.09& 893.52\\
 23& 2002.204& 2334.05&& 2000.958& 1122.41&  708.56  && 2002.204&1383.22& 950.83\\
 24& 2014.455& 1628.64&& 2011.707&  686.55&  218.21  && 2014.455&1211.00& 417.64\\
\noalign{\smallskip}
\noalign{\smallskip}
 &Year & $A_{\rm m}$ &&Year & $A_{\rm m}$ &$ A^*_{\rm S}$ && Year & $A_{\rm m}$&$ A^*_{\rm N}$ \\
\noalign{\smallskip}
 12& 1878.958&   16.22&& 1879.042&    9.58& 7.44  && 1878.707&  0.28& 17.79\\
 13& 1888.874&   63.88&& 1889.455&    4.32& 72.31 && 1890.204& 30.35& 33.95\\
 14& 1901.455&   25.71&& 1900.791&   13.86& 31.82 && 1901.455&  6.10& 19.61\\
 15& 1913.623&    7.63&& 1912.373&    0.56& 33.20 && 1913.538&  2.78&  5.17\\
 16& 1923.623&   48.63&& 1923.623&   28.27& 20.36 && 1923.623& 20.36& 28.27\\
 17& 1933.707&   32.31&& 1933.790&   18.13& 29.00 && 1933.538&  1.63& 85.90\\
 18& 1944.124&  105.32&& 1944.373&   40.91& 64.55 && 1943.958& 34.42& 84.41\\
 19& 1954.288&   24.14&& 1954.288&    4.44& 19.71 && 1954.288& 19.71&  4.44\\
 20& 1964.791&   49.45&& 1964.456&   42.42& 12.65 && 1964.874&  3.39& 61.96\\
 21& 1976.874&  133.95&& 1976.874&   76.43& 57.52 && 1975.455& 41.90& 131.06\\
 22& 1986.707&   91.77&& 1985.958&   42.60& 89.46 && 1986.707& 29.35& 62.41\\
 23& 1996.624&   83.90&& 1996.791&   23.52& 66.40 && 1995.958& 39.82& 51.20\\
 24& 2008.874&    5.93&& 2008.124&    1.89& 30.75 && 2008.958&  2.01&  4.57\\
 \noalign{\smallskip}
\hline
\end{tabular}
\label{table2}
}
\end{table}

\begin{table}
{\scriptsize
\caption[]{The rise time (RT), decline time (DT), and length (L) [years]
of Sunspot (ISSN) Cycles~12\,--\,14 determined from 13-month 
smoothed monthly mean ISSN and those of  
 WSGA, NSGA, and SSGA Cycles 12\,--\,14 determined from the
 corresponding time series of 13-month smoothed monthly
 mean areas of the sunspot groups in whole sphere,
 northern hemisphere, and southern hemisphere, respectively.}
\begin{tabular}{lccccccccccccccc}
\noalign{\smallskip}
\hline
Cycle &\multicolumn{3}{c}{ISSN}&& \multicolumn{3}{c}{WSGA} && 
\multicolumn{3}{c}{NSGA}  &&\multicolumn{3}{c}{SSGA}\\
  \noalign{\smallskip}
\cline{2-4}
\cline{6-8}
\cline{10-12}
\cline{14-16}
\noalign{\smallskip}
 &RT&DT&L&& RT&DT&L&& RT&DT&L&& RT&DT&L\\
\noalign{\smallskip}
\hline
\noalign{\smallskip}
12&  5.00 & 6.17& 11.17&&  5.00&  4.92&  9.92&&  3.33&  7.08& 10.41&&  5.25&  6.25& 11.50\\
13&  3.92 & 8.00& 11.92&&  5.17&  7.41& 12.58&&  4.67&  6.67& 11.34&&  3.75&  7.50& 11.25\\
14&  4.08 & 7.33& 11.41&&  4.00&  8.17& 12.17&&  5.25&  6.33& 11.58&&  6.00&  6.08& 12.08\\
15&  4.17 & 5.91& 10.08&&  3.91&  6.09& 10.00&&  5.33&  5.92& 11.25&&  4.09&  6.00& 10.09\\
16&  4.75 & 5.42& 10.17&&  2.66&  7.42& 10.08&&  2.66&  7.50& 10.17&&  4.58&  5.33&  9.91\\
17&  3.58 & 6.84& 10.42&&  3.58&  6.84& 10.42&&  3.75&  6.84& 10.58&&  5.09&  5.33& 10.42\\
18&  3.25 & 6.92& 10.16&&  3.25&  6.92& 10.16&&  5.25&  4.66&  9.91&&  3.50&  6.83& 10.33\\
19&  3.92 & 6.59& 10.50&&  3.67&  6.83& 10.50&&  5.25&  4.92& 10.17&&  3.50&  7.08& 10.59\\
20&  4.08 & 7.33& 11.42&&  5.75&  6.34& 12.08&&  3.17&  9.25& 12.42&&  5.00&  5.58& 10.58\\
21&  3.75 & 6.75& 10.50&&  4.83&  5.00&  9.83&&  2.50&  6.59&  9.08&&  6.50&  4.75& 11.25\\
22&  2.83 & 6.84&  9.67&&  2.75&  7.17&  9.92&&  3.50&  7.34& 10.83&&  4.83&  4.42&  9.25\\
23&  3.92 & 8.67& 12.58&&  5.58&  6.67& 12.25&&  4.17&  7.17& 11.33&&  6.25&  6.75& 13.00\\
24&5.33&&&&           5.58&&&&           3.58&&&&           5.50\\
\noalign{\smallskip}
\hline
\end{tabular}
\label{table3}
}
\end{table}

\begin{SCtable}
{\scriptsize
\caption[]{The difference between the epochs [years]. 
$\delta T_{\rm Mw}$: $T_{\rm M}$ of ISSN cycle minus epoch of  
 $A_{\rm M}$ of WSGA cycle,   
$\delta T_{\rm Mns}$: epoch of $A_{\rm M}$ NSGA cycle 
 minus epoch of $A_{\rm M}$ of SSGA cycle,      
$\delta T_{\rm mw}$: $T_{\rm m}$ of ISSN cycle minus epoch of $A_{\rm m}$ 
of WSGA cycle, and   
$\delta T_{\rm mns}$: epoch  of $A_{\rm m}$ NSGA cycle  minus 
 epoch of $A_{\rm m}$ of SSGA cycle.      
Only those   values  
 correspond to the revised sunspot number (SN) series  
 differ with the values correspond to the original time series of
 $R_{\rm Z}$ are given in parentheses. 
}
\begin{tabular}{lccccc}
\noalign{\smallskip}
\hline
Cycle& $\delta T_{\rm Mw}$& $\delta T_{\rm Mns}$& $\delta T_{\rm mw}$& 
$\delta T_{\rm mns}$&\\
\noalign{\smallskip}
\hline
\noalign{\smallskip}
 12&    0.00&   -1.59&    0.00&    0.33\\
 13&    0.00&    0.17&    1.25&   -0.75\\
   & &&(1.33)&\\ 
 14&    0.67&   -1.41&    0.59&   -0.66\\
 15&    0.09&    0.08&   -0.17&   -1.16\\
 16&    2.00&   -1.92&   -0.09&    0.00\\
 17&    0.00&   -1.09&    0.00&    0.25\\
 18&    0.00&    2.17&    0.00&    0.42\\
 19&    0.25&    1.75&    0.00&    0.00\\
 20&   -1.66&   -2.25&    0.00&   -0.42\\
 21&   -1.75&   -2.59&   -0.67&    1.42\\
 22&    0.08&   -2.08&    0.00&   -0.75\\
   &(0.42) &&&\\ 
 23&   -1.91&   -1.25&   -0.25&    0.83\\
   &(-0.33) &&(0.00)&\\ 
 24&   -0.17&   -2.75&    0.08&   -0.83\\
 \noalign{\smallskip}
\hline
\end{tabular}
\label{table4}
}
\end{SCtable}

 There seems to be a difference in 
the long-term trends in  maximum and minimum values 
 of the solar cycles. In fact, 
  no significant correlation is found between maximum and
 minimum values ($r = 0.5$, less than 95\,\% confidence level, 
$i.e.$ $P > 0.05$). There are trends of  about five-solar 
 cycles time scale (44\,--\,55-year periodicity)
in the variation of  maxima. In the case of minimum, a periodicity seem to
 evolve from 33 years to 66 years with an increase in amplitude. All of these
 results 
 hold good also for the  corresponding variations in $A^*_{\rm M}$ and  
$A^*_{\rm m}$. 

The North--South asymmetry of a solar-activity index  is usually
determined as $\frac{N - S}{N + S}$,
where $N$ and $S$ are the activity indices in the northern
and southern solar hemispheres~(disadvantages of some
 other definitions are  discussed by \opencite{jg97a}; \opencite{jj16}).
 Figure~4  shows the 
 variations in the North--South asymmetry of $A_{\rm M}$,  $A^*_{\rm M}$, 
 $A_{\rm m}$, and  $A^*_{\rm m}$ (Note: in the case of $A_{\rm M}$ and 
 $A_{\rm m}$ in many cycles the northern and southern epochs are not  same).
As can be seen in Figure~4(a), the patterns of variations  in the
 North--South 
asymmetry of $A_{\rm M}$ and  $A^*_{\rm M}$ are closely similar 
(the corresponding correlation coefficient $r = 0.90$,
 significant at the 99.99\,\% confidence 
level found from Student's t-test, $P< 0.001$). 
There are suggestions that 
in Cycles 12, 13, and 24 the $A_{\rm M}$ of the southern hemisphere is 
 considerably larger than that of the northern hemisphere. Overall 
 there seems to be  a trend of a 130\,--\,140-year cycle  in the North--South 
asymmetry of $A_{\rm M}$ (and of  $A^*_{\rm M}$).  However, the 143 year 
data used here are not adequate to obtain the accurate value of this 
periodicity.  
On the other hand, in Cycle~18 the asymmetry of $A_{\rm M}$
has   a  low value. Hence, 
there are also  indications on   the  trends of about 55-year cycles. 
As can be seen in Figure~4(b), in all cycles except Cycles~12 and 24 
 the values   of the North--South 
asymmetry in $A_{\rm m}$ and  $A^*_{\rm m}$ are almost the same 
(the corresponding 
$r = 0.844$, significant at the 99.9\,\% confidence level). 

 The long-term periodicities of the North--South asymmetry in 
  minimum   seem to be  different from those of maximum.        
There is a strong suggestion that in Cycles 12, 17, and 20  
  $A_{\rm m }$ of the  northern hemisphere is significantly larger than 
that of the southern hemisphere,  and 
in Cycles 13, 15, 19, and to some extent in Cycle~23,  $A_{\rm m }$
 of the southern hemisphere is significantly larger than that of 
 the northern hemisphere. The overall
 pattern suggests the
 presence of  about 44\,--\,66-year cycles in the North--South asymmetry
 of $A_{\rm m}$.  A similar conclusion can be also  drawn for   
 the North--South asymmetry in  $A^*_{\rm m}$.
 Variation in the
 North--South asymmetry of minimum  might have mainly
 contributed to   a 50\,--\,60-year peak that was found dominant in the power
  spectra of the North--South asymmetry in sunspot 
activity~(see Figure~7 of \opencite{jg97a}). 

There is no 
significant correlation  
between maximum   of solar cycle ($R_{\rm M}$  or $A_{\rm M}$ of WSGA cycle)
and its North--South asymmetry. There is also no 
significant correlation   
between minimum of solar cycle ($R_{\rm m}$ or  $A_{\rm m}$ of WSGA cycle)
and its North--South asymmetry (maximum value of $|r| < 0.3$).

\begin{figure}
\centering
\includegraphics[width=\textwidth]{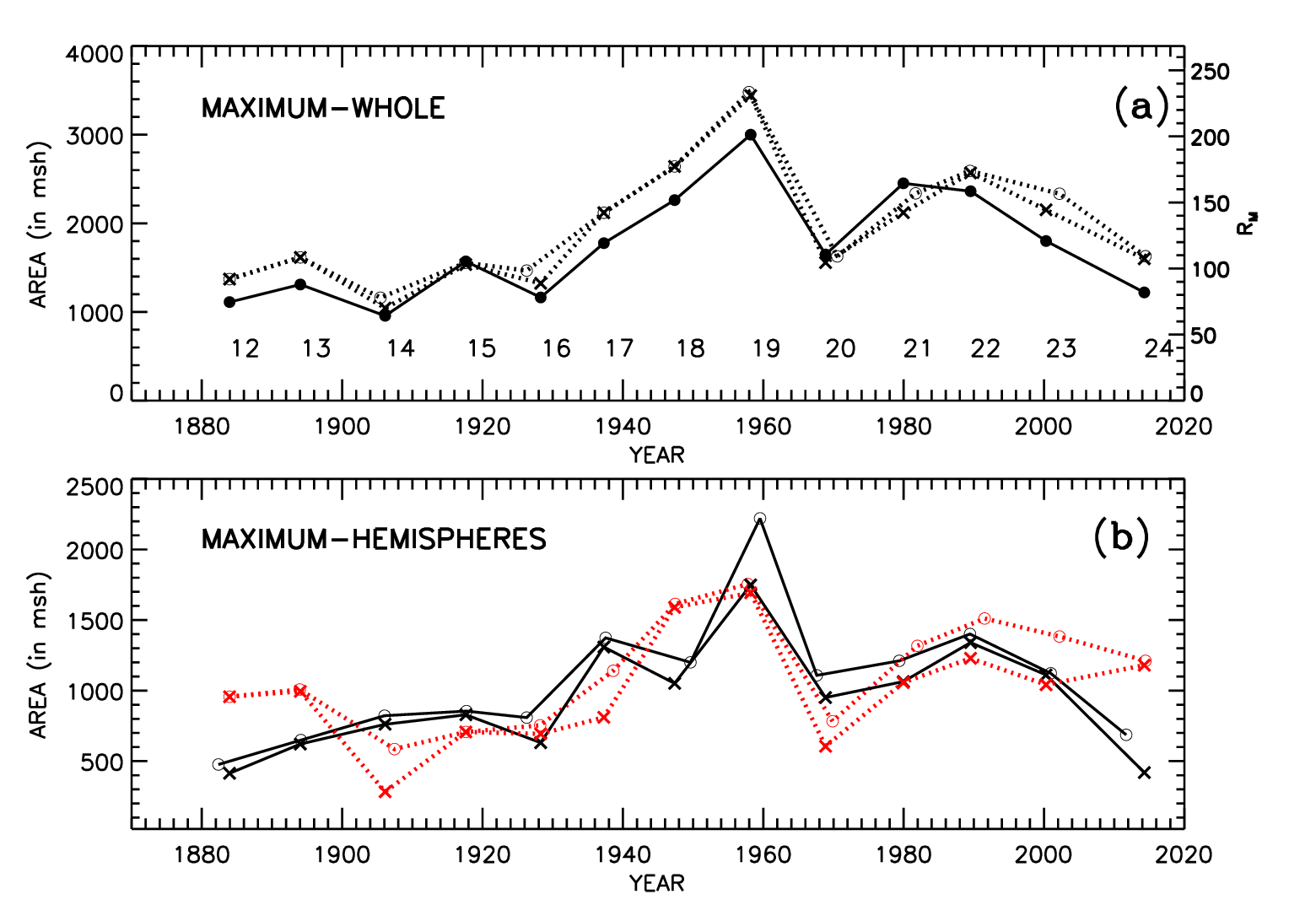}
\caption{The values of maxima of  
Solar Cycles 12\,--\,24 {\it versus} their corresponding epochs.  
The {\it crosses} indicate the values 
 of $A^*_{\rm M}$,  which  correspond to the
 epochs of $R_{\rm M}$,   the {\it open-circles} indicate 
the values of $A_{\rm M}$,  which correspond to the 
epochs of the actual maxima,  and   
the {\it filled-circles} indicate the values
 of $R_{\rm M}$.
({\bf a}) The {\it dotted curves} represent the variations in $A_{\rm M}$  and 
 $A^*_{\rm M}$ obtained from the whole sphere sunspot group data (WSGA)
 and the {\it continuous curve} represents the variation in   $R_{\rm M}$
 ({\bf b}) The {\it continuous curves} represent the variations in $A_{\rm M}$ 
 and  $A^*_{\rm M}$  obtained from the northern hemisphere data (NSGA), and
 the {\it dotted curves} (red) represent the corresponding variations 
obtained from the southern-hemisphere data (SSGA),
In panel {\bf a }the Waldmeier numbers of  solar cycles are also given.}
\label{f2}
\end{figure}

\begin{figure}
\centering
\includegraphics[width=\textwidth]{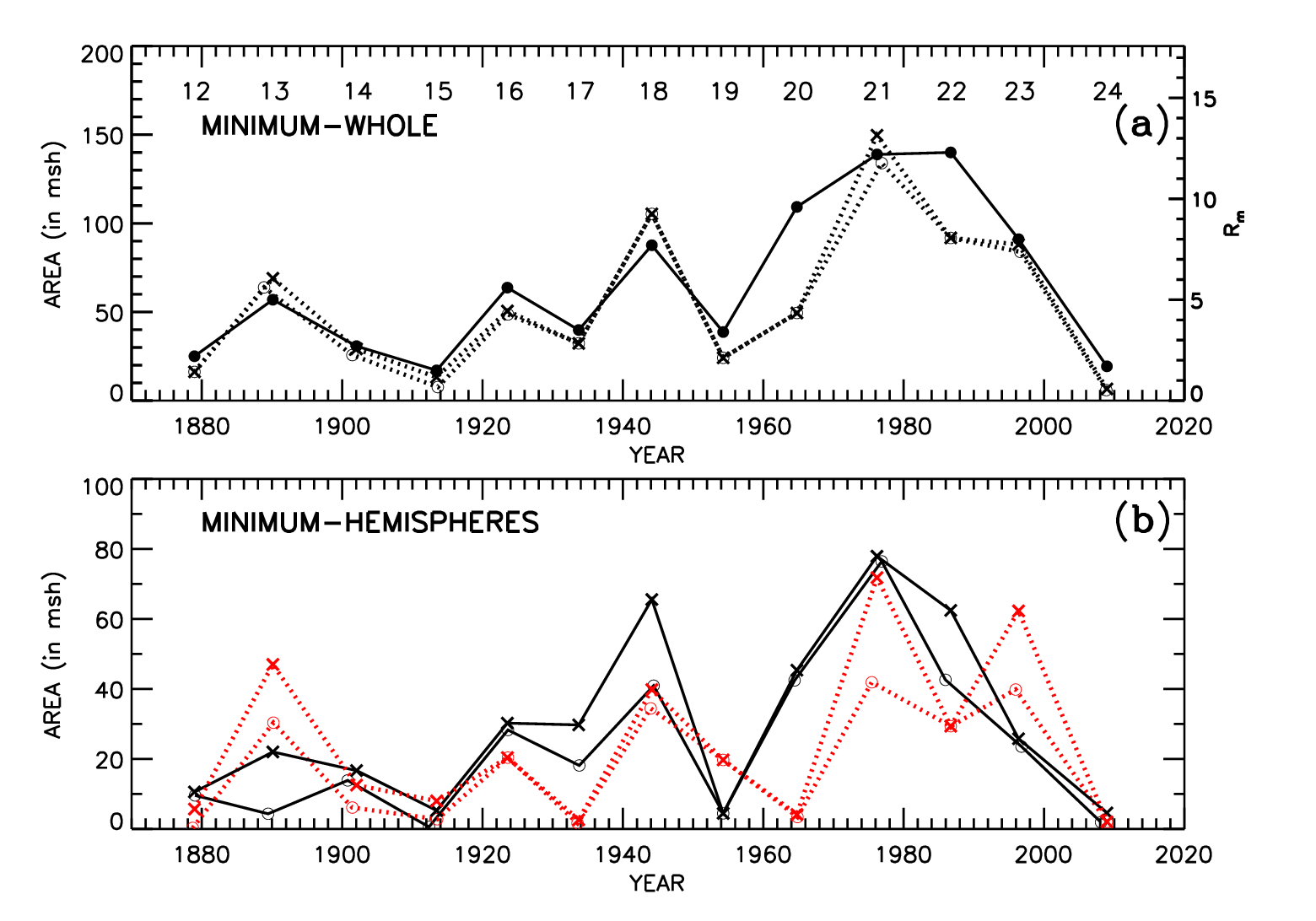}
\caption{The values of the  minima  of the 
Solar Cycles 12\,--\,24 {\it versus} their corresponding epochs.  
The {\it crosses} indicate the values 
 of $A^*_{\rm m}$ that  correspond to the
 epochs of $R_{\rm m}$, the {\it open-circles} indicate 
the values of $A_{\rm m}$ that correspond to the 
epochs of the actual  minima, and   
the {\it filled-circles} indicate the values
 of $R_{\rm m}$.
({\bf a}) The {\it dotted curves} represent the variations in $A_{\rm m}$  and 
 $A^*_{\rm m}$ obtained from the whole-sphere sunspot-group data (WSGA)
 and the {\it continuous curve} represents the variation in   $R_{\rm m}$, and 
 ({\bf b}) The {\it continuous curves} represent the variations in $A_{\rm m}$ 
 and  $A^*_{\rm m}$    obtained from the northern hemisphere data (NSGA) and 
the {\it dotted curves} (red) represent the corresponding variations obtained from 
the southern hemisphere data (SSGA). 
In panel {\bf a} the Waldmeier numbers of  solar cycles are also given.}
\label{f3}
\end{figure}

\begin{SCfigure}
\centering
\vspace{1.9cm}
\includegraphics[width=7.5cm]{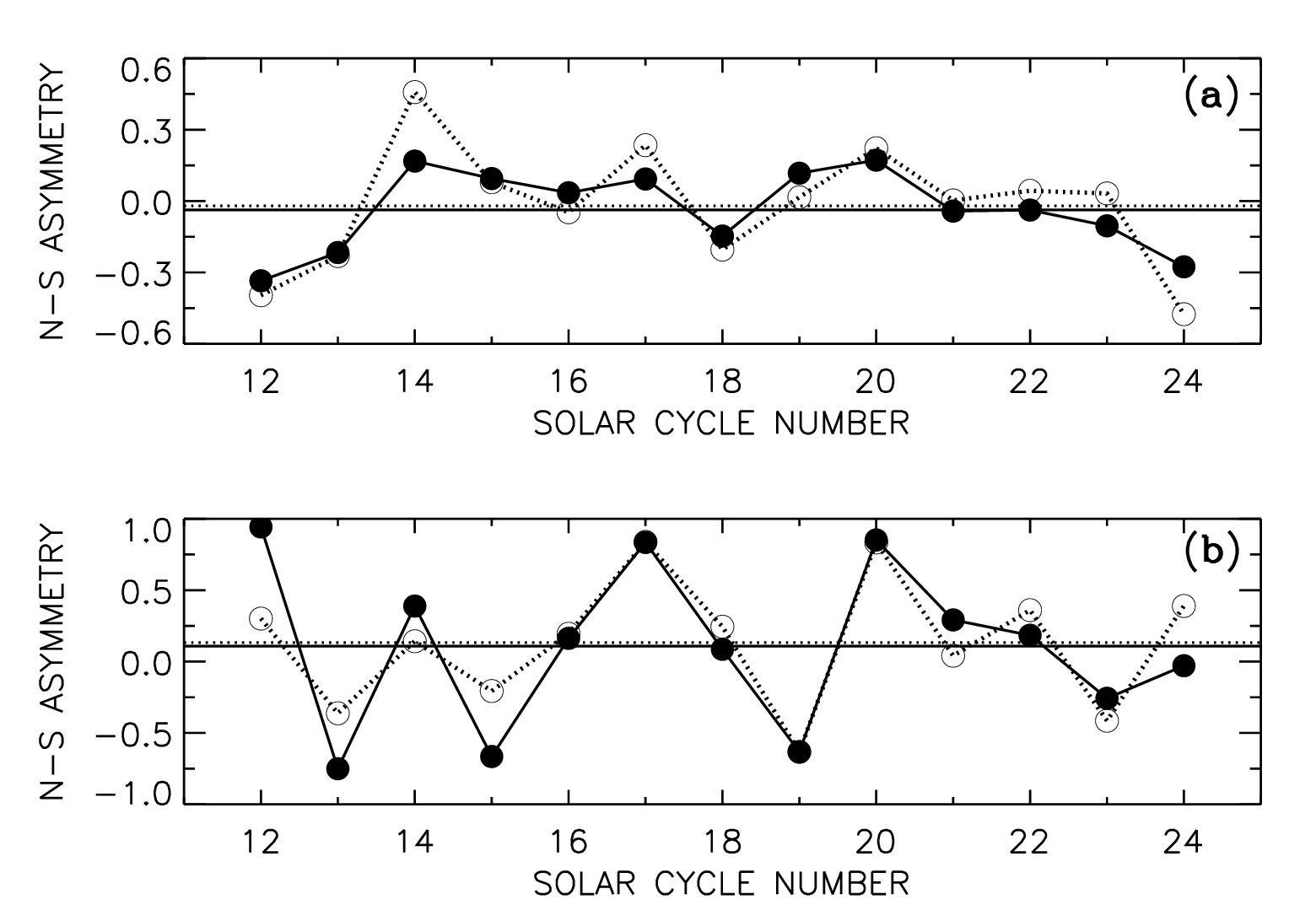}
\caption{The cycle-to-cycle variations in the North--South asymmetry
 ({\bf a}) in the maximum and ({\bf b}) in the minimum of the
 Solar Cycles 12\,--\,24.
 The {\it open-circle-dotted curves} represent the North--South asymmetry in  
 $A^*_{\rm M}$ and   $A^*_{\rm m}$ 
and the  {\it filled circle-continuous curves}  
represent the North-South asymmetry in 
 $A_{\rm M}$ and  $A_{\rm m}$.
The positive
 and negative values imply the North dominance and the South dominance,
respectively.}
\label{f4}
\end{SCfigure}

\begin{figure}
\centering
\includegraphics[width=\textwidth]{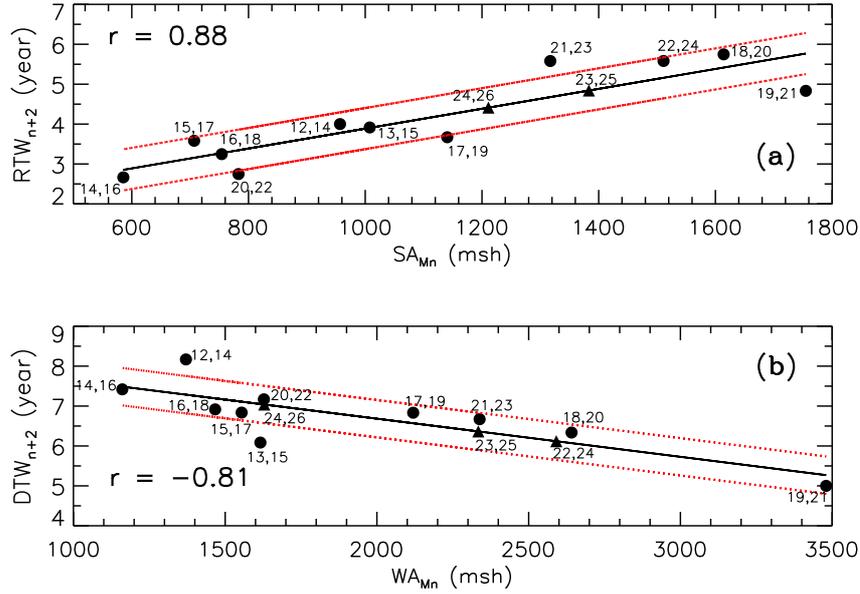}
\caption{Scatter plots of ({\bf a}) $SA_{{\rm M}_n}$: 
the maximum [$A_{\rm M}$] of
 Solar Cycle~$n$ of the area of sunspot groups  in 
 southern hemisphere (SSGA)
{\it versus} $RTW_{n+2}$: the rise time (RT) of Cycle~$n+2$  of the   
area of sunspot groups in whole sphere (WSGA),  
 ({\bf b})  $WA_{{\rm M}_n}$:  $A_{\rm M}$ of Solar Cycle~$n$
of the area of sunspot groups  in whole hemisphere (WSGA)
{\it versus}  $DTW_{n+2}$: the decline time (DT) of Cycle~$n+2$
of the   area of sunspot groups in whole sphere (WSGA).
The {\it continuous line} 
represents the best linear least-square fit and the {\it dotted lines} (red) 
are drawn at one-{\it rmsd} ({\it root-mean-square deviation}). The
values of the correlation coefficient [$r$] are also given.
The {\it filled triangles} represent the predicted values of the Cycles~25 
 and 26 in the case of ({\bf a})  and the Cycles~24, 25, and  26 in the
 case of ({\bf b}).}
\label{f5}
\end{figure}

\begin{figure}
\centering
\includegraphics[width=\textwidth]{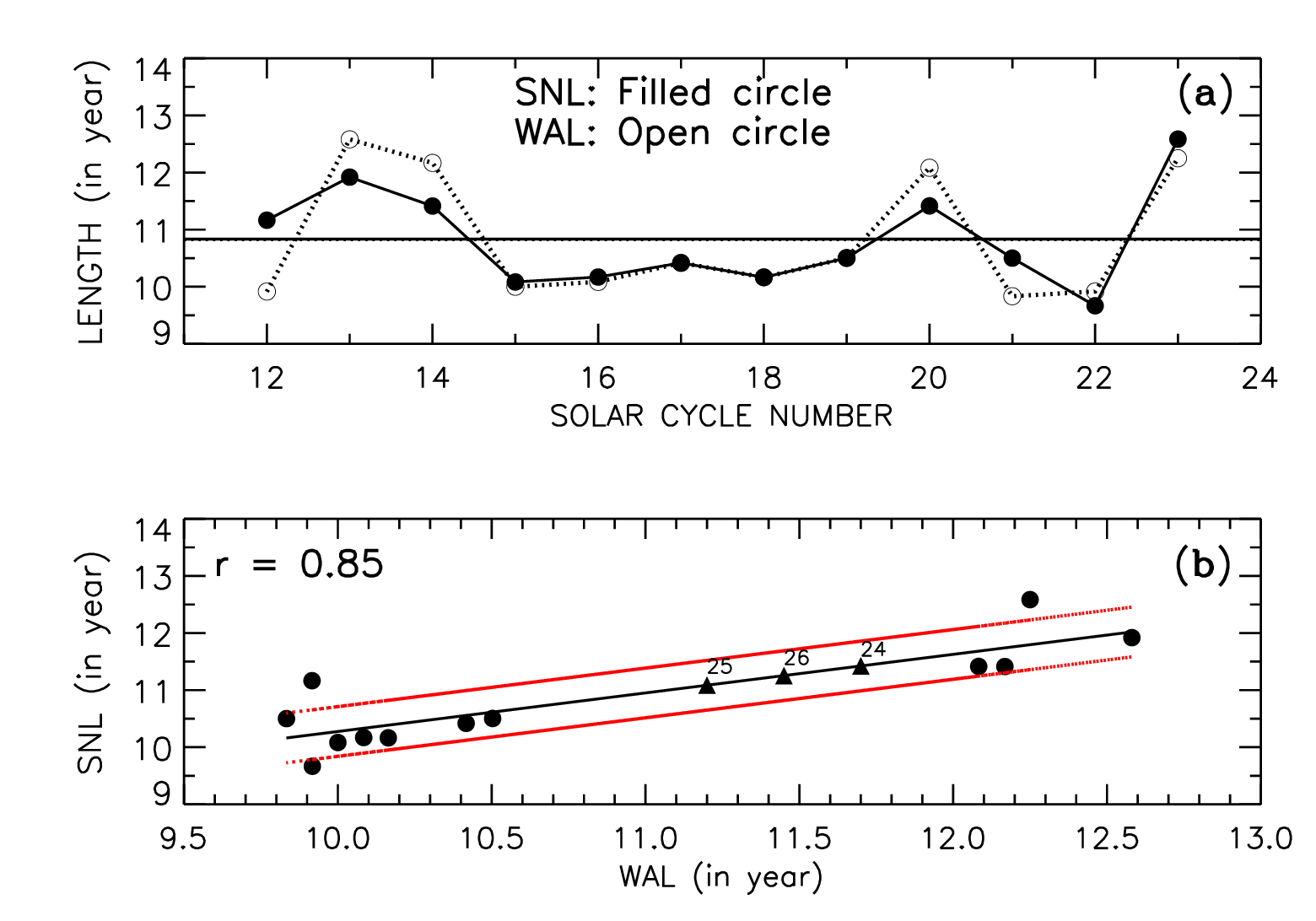}
\caption{({\bf a})  NSL and WSL, $i.e.$ lengths of ISSN  and WSGA cycles,
 respectively, {\it versus} solar-cycle number. ({\bf b}) Scatter 
plot of NSL {\it versus} WAL.  
The {\it continuous line}
represents the best linear least-square fit and the {\it dotted lines} (red)  
are drawn at one-{\it rmsd}. The
value of $r$ is also given.
The filled triangles represent the predicted values of the cycles 24, 25,
 and 26.}
\label{f6}
\end{figure}

\subsection{Prediction of Lengths of Upcoming Solar Cycles}

Using the values given in Tables~2 and 3  we determined correlations 
among  the several parameters of solar cycles of different indices. 
However, only the relations 
described below are found to be reasonably statistically significant.

Figure~5(a) shows the  relationship between  the maximum [$A_{\rm M}$] of 
 the SSGA Cycle~$n$  (namely $SA_{{\rm M}_n}$)  and the rise time (RT) of the
 WSGA Cycle~$n+2$ (namely $RTW_{n+2}$).  Figure~5(b) shows the relationship
 between  the maximum   [$A_{\rm M}$] of the WSGA Cycle~$n$ 
 (namely $WA_{{\rm M}_n}$)
 and the decline-time (DT) of the WSGA Cycle~$n+2$ (namely $DTW_{n+2}$).
 We obtained the following linear relations:
$$RTW_{n+2} = (0.0025\pm0.00045) SA_{{\rm M}_n} + 1.39\pm0.53    \eqno(1)$$ 
\noindent and 
$$DTW_{n+2} = (-0.00096\pm0.00024) WA_{{\rm M}_n} + 8.6\pm 0.50. \eqno(2)$$ 

The fit of  Equation~1 to the data is  good.  The slope is 5.55 times 
greater than the corresponding standard deviation  and except for the  
two data points (19, 20) and (21, 23), most of the remaining nine 
data points are within or very close to the one-root-mean-square  
deviation ($rmsd = 0.51$).
The corresponding correlation coefficient
$r = 0.88$ is statistically significant at the 99.9\,\% confidence level 
(Student's  $t = 5.5$, which 
is much higher than the value 4.78 $t$-significance at the 0.1\,\% level,
 $P < 0.001$,  for nine degrees of freedom) and
 $\chi^2 = 2.29$  is much lower than the value 18.307 $\chi^2$ significant
on 5\% level, $P = 0.05$, for ten degrees of freedom. 
(Note: we have also obtained a positive correlation between 
$WA_{{\rm M}_n}$  and $RTW_{n+2}$, but it is found  considerably 
 smaller than that of $SA_{{\rm M}_n}$ and $RTW_{n+2}$.) 

The fit of  Equation~2 to the data is also reasonably good (slightly weaker 
than that of Equation~1). The  slope  is four times 
greater than the corresponding standard deviation 
  and except the  
two data points (12, 14) and (13, 15), most of the remaining eight
data points are  within or vary close to the one-{\it rmsd}  level.
The corresponding correlation coefficient
$r = - 0.81$ is statistically significant at the 99\,\% confidence level 
($t = 3.9 $, which is slightly higher than the value 3.36 $t$-significant
 at the 1\,\% level, $P =0.01$,   for 8 degrees of freedom) and
 $\chi^2 = 3.1$  is much lower than the 
value 16.919 $\chi^2$-significant  5\,\% level, $P = 0.05$, for 9
 degrees of freedom.

By substituting the values 1383.22 msh and 1211.0 msh of 
$SA_{{\rm M}_n}$ of the Cycles~23 and 24 in Equation~1 we 
obtained the values $4.84 \pm 0.51$ years and $4.41 \pm 0.51$ years  for 
  $RTW_{n+2}$, $i.e.$ the rise times of the  WSGA Cycles~25 and 26,
 respectively. 
By substituting  the values 2591.13 msh, 2334.05 msh, and 1628.64 msh of
$WA_{{\rm M}_n}$ of the WSGA Cycles~22, 23, and 24 in Equation~2 we
obtained the values $6.12 \pm 0.47$ years, $6.36 \pm 0.47$ years, and
$7.04 \pm 0.47$ years  of   
  $DTW_{n+2}$, $i.e.$ the decline-times of WSGA Cycles~24, 25, and 26,
 respectively. All of these predicted values are also shown in Figure~5.
 The sum of the values of the rise- and  the decline-times 
gives the lengths (the values of L), $11.7 \pm 0.15$ years, 
$11.2 \pm 0.2$ years, and $11.45 \pm 0.3$ years
   of the WSGA Cycles~24, 25, and 26
(in the case of the WSGA Cycle~24 the observed value of the rise time given in
Table~3  is used). These predicted values of the  
lengths  suggest that  $2020.57 \pm 0.15$  (July 2020), 
$2031.77 \pm 0.2$  (October 2031), and $2043.22 \pm 0.3$
  (March 2043) would be the 
 minimum epochs (start dates) of WSGA  
Cycles~25, 26, and 27, respectively,  and  the predicted rise times 
suggest that $2025.41 \pm 0.51$   (May 2025) and $2036.18 \pm 0.51$ 
 (March 2036)  would be  the maximum
 epochs of WSGA Cycles 25 and 26, respectively.

Figure~6(a) shows the variations in the lengths of  ISSN cycle
 (namely SNL) and  WSGA cycle (namely WAL). As can be seen in this
 figure, only in a
 few places a considerable difference exists between SNL and WAL. However, 
 both SNL and WAL vary almost identically. In an earlier analysis  a cosine fit 
to the data (length values) of ISSN Cycles 6\,--\,22 has suggested the 
existence of a eight-cycle periodicity in the  length of ISSN 
cycle~(see Figure~5 of \opencite{jbu05}). Obviously, the patterns of the 
 variations in SNL and WAL are consistent with that earlier result. 
Figure~6(b) shows the relationship between WAL and SNL.
 There exists a considerable correlation 
between WAL and  SNL ($r = 0.85$, the corresponding $P < 0.001$).
  We obtained the following  linear relation between WAL and SNL: 
$$SNL = (0.675\pm0.132) WAL + 3.52\pm1.43 \quad .    \eqno(3)$$
The corresponding  least-square fit  is reasonably good ($\chi^2 = 3.0$ is 
 small, the slope is 5.1 times larger than the corresponding standard 
deviation). The $rmsd  = 0.43$  is reasonably small. 
 Except for two data points (that correspond to the Cycles 12
 and 23), all of the remaining data points are inside or on the lines of 
 one-{\it rmsd}.

 By substituting  the  predicted values of WAL in Equation~3 
 we obtained the values $11.42\pm 0.43$ years, $11.09\pm 0.43$ years, 
and $11.25\pm 0.43$ years for the lengths of ISSN
 Cycles~24, 25, and 26, respectively.
This suggests that the minimum epochs (start dates) of ISSN Cycles 25, 
26, and 27 would be $2020.38\pm 0.43$  (May 2020), 
$2031.47 \pm 0.43$ (June 2031), and
 $2042.72\pm 0.43$  (September 2042),
 respectively. All of these values seem to be  
slightly less than the 
the corresponding predicted values of the WSGA Cycles 24, 25, and 26.  
However, the differences between the predicted values of the 
ISSN and WSGA cycles  are not significant.
 
When we have used lengths of Cycles 12\,--\,13 of the  revised time series 
of sunspot numbers (SN) the correlation 
between the WAL and NAL is found to be 0.86 ($\chi^2 = 2.84$)  and  
 $11.4\pm 0.4$ years, $11.07\pm 0.4$ years,
and $11.23\pm 0.4$ years are found for the lengths of Cycles~24, 25, and 26,
 respectively.

\inlinecite{dgt08} found that the Waldmeier effect is not present  in the 
cycles of  sunspot-group area. Here, we have also found the same. 
 That is, we find that there exists no significant correlation between  
rise time and amplitude of  WSGA, NSGA, and SSGA cycles.
 Also no significant correlation is found 
between the rise times of ISSN and  area cycles. Hence, here we cannot
 predict the rise times of the ISSN Cycles~25 
and 26 by using the above-predicted rise times of WSGA Cycles~25 and 26.
 
As per the above prediction,  the length of Solar Cycle~25 would be smaller 
than those of Cycles 24 and 26. Hence, as per the well-known inverse 
relationship between cycle amplitude and length, the above predicted lengths 
indicate that Cycle 25 would be stronger than Cycles 24 and 26. 
This contradicts our prediction~(\opencite{jj15}, \citeyear{jj17}).  However,  
 the variations in solar 
cycle length and amplitude are not perfectly in phase or anti-phase
(\opencite{hs97}). In fact, only a weak  correlation exists 
 between length and amplitude of a cycle~($e.g.$ \opencite{solanki02}). 

There exits a highly statistically  significant inverse 
relationship  between length of a cycle and the amplitude of the 
following cycle (\opencite{hath94}; \opencite{solanki02}; \opencite{hath15}).
As per this relationship  the above-predicted  lengths of the  ISSN-cycles
 24, 25, and 26  indicate  that 
Cycle 25 would be stronger than  the Cycle 24 (Note: the length, 12.25 year, of Cycle~23 
is larger than the predicted length of Cycle~24).
 Hence, this    is also a contradiction to
 our earlier prediction that Cycle~25 will be weaker than
 Cycle~24~(\opencite{jj15}, \citeyear{jj17}).  On the other hand 
the corresponding correlations of all of the present predictions are not
very high.

\subsection{Asymmetry at the Epochs of the Actual Maxima  and Minima 
of NSGA- and SSGA-cycles}
It should be noted that the epochs of the actual maximum [$A_{\rm M}$], 
as well as  those of  the actual  minimum [$A_{\rm m}$],  of NSGA- and 
SSGA-cycles are different in  many cycles. 
In Section 3.1 we have shown the   
asymmetry between the $A_{\rm M}$ of NSGA and $A_{\rm M}$ of SSGA and 
 between the $A_{\rm m}$ of NSGA and $A_{\rm m}$ of SSGA 
(the filled-circle-continuous curves in Figure~4).

The upper panel of Figure~7 shows the cycle-to-cycle modulation 
in $A^*_{\rm N}$, $i.e.$ 
the  value of  NSGA  at the epoch of $A_{\rm M}$  of SSGA-cycle, 
and the modulation in  $A^*_{\rm S}$, $i.e.$ the 
value of  SSGA  at the epoch of $A_{\rm M}$  of NSGA-cycle
 (the values $A^*_{\rm N}$ and $A^*_{\rm S}$ are also given in Table~2). 
In this figure the modulation  in  $A_{\rm M}$ of NSGA-cycle 
and that in  $A_{\rm M}$ of SSGA-cycle (which are already 
shown in Figure~2) are also shown  for the sake of comparison of these with 
those of $A^*_{\rm N}$ and $A^*_{\rm S}$. In the lower panel of Figure~7 
 the cycle-to-cycle modulation in the  asymmetry
 between  the $A_{\rm M}$ of NSGA and $A^*_{\rm S}$ 
and  the  modulation in the asymmetry  between the $A_{\rm M}$ of SSGA
 and $A^*_{\rm N}$  are  shown. Figure~8 is same as Figure~7, but for the 
data at the epochs of actual minima of  NSGA- and SSGA-cycles. 

\begin{figure}
\centering
\includegraphics[width=\textwidth]{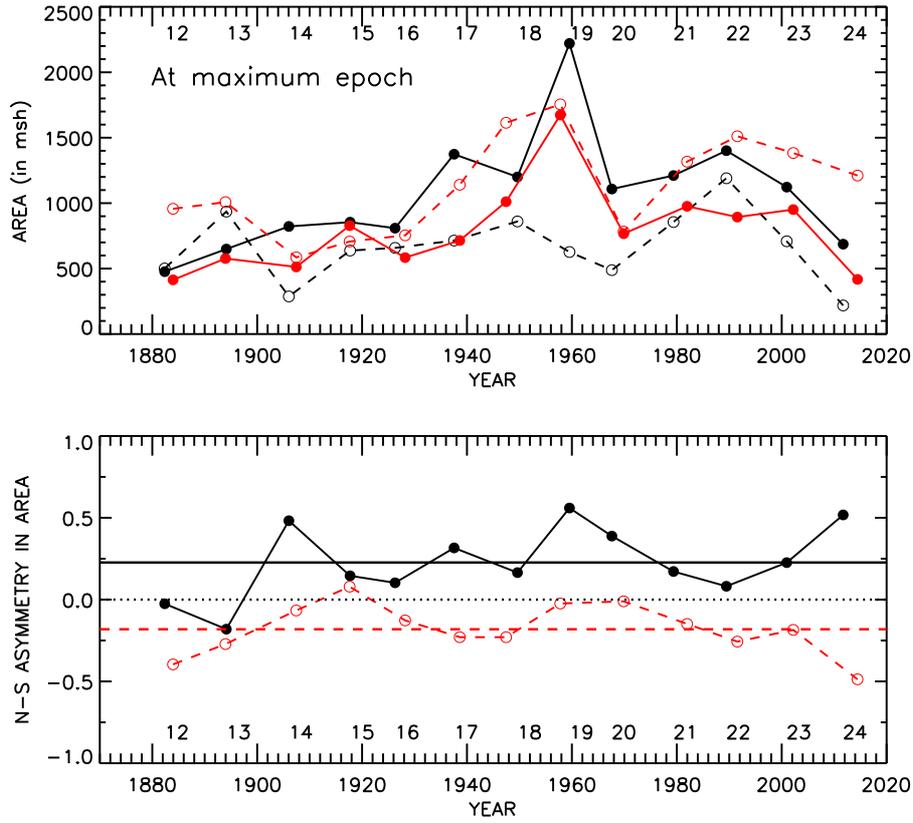}
\caption{{\bf Upper panel:} The maximum [$A_{\rm M}$]
value of NSGA (SSGA) cycle (also 
shown in Figure~2) and the value of $A^*_{\rm S}$ ($A^*_{\rm N}$)
of  SSGA (NSGA) at the epoch of the maximum of NSGA (SSGA) 
cycle (which are also given Table~2) {\it versus} time ($i.e.$ epoch
 of maximum). 
The {\it filled circle-continuous curves}
 represent  the modulations in  $A_{\rm M}$ of NSGA-cycle and $A^*_{\rm N}$, 
$i.e.$ the  value of  NSGA  at the epoch of the maximum of the SSGA-cycle.   
 The {\it open-circle-dashed curves} represent the variations 
in $A_{\rm M}$ of  SSGA-cycle and $A^*_{\rm S}$, 
$i.e.$ the  value of  SSGA  at the epoch of the maximum of the NSGA-cycle.
The corresponding quantities at the epochs of maxima of NSGA- and SSGA-cycles   
 are shown by  {\it black} and {\it red} colors, respectively. 
{\bf Lower panel:} The {\it black filled-circle-continuous curve}  and 
the {\it red open-circle-dashed curve} represent the variations of
 North--South asymmetry in the corresponding quantities
  at the epochs of the maxima of NSGA- and SSGA-cycles, respectively. 
The positive  and negative values imply the North dominance and the
 South dominance, respectively.}
\label{f7}
\end{figure}

\begin{figure}
\centering
\includegraphics[width=\textwidth]{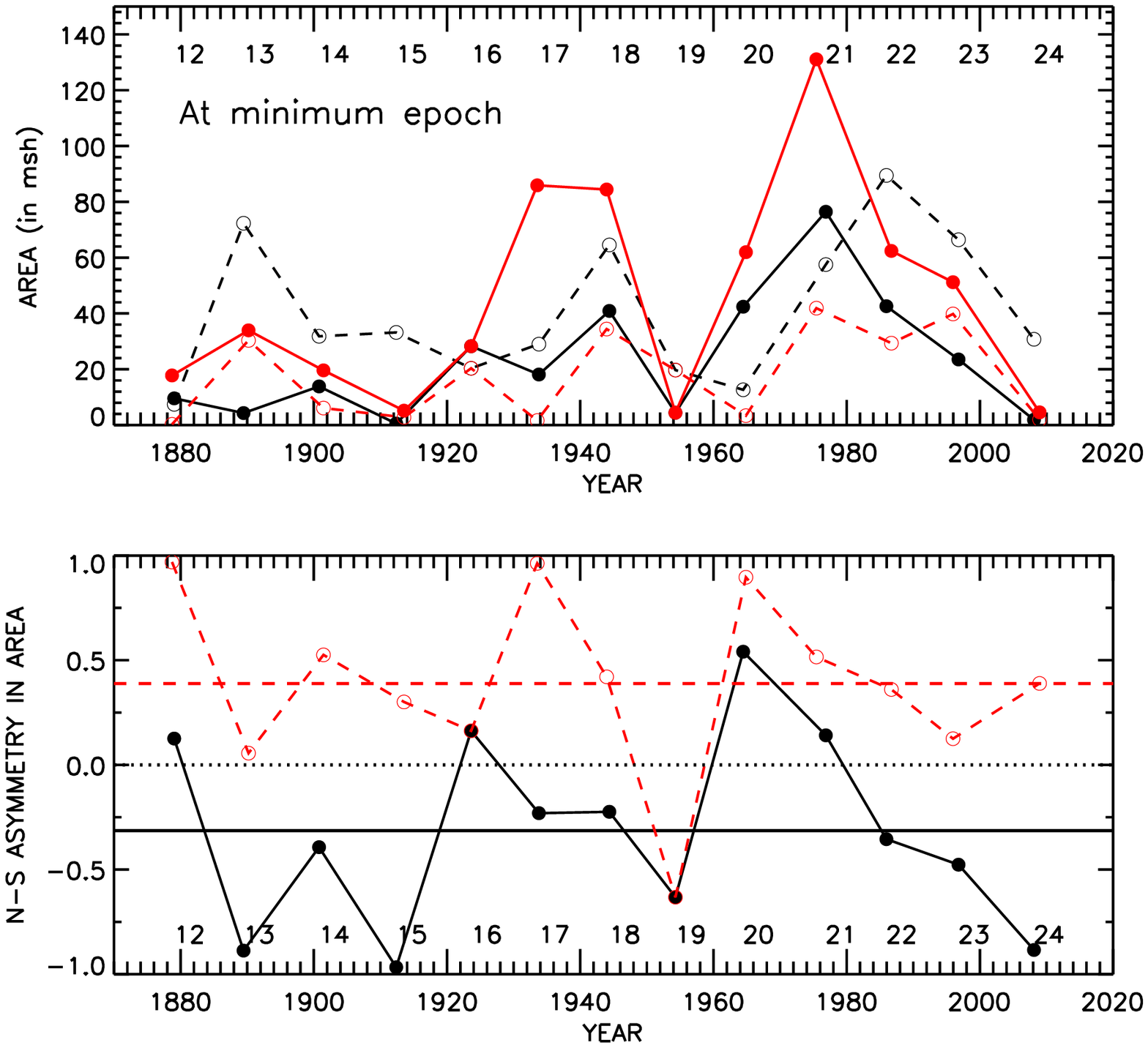}
\caption{{\bf Upper panel:} The minmum [$A_{\rm m}$] 
value of NSGA (SSGA) cycle (also 
shown in Figure~3) and the value of $A^*_{\rm S}$ ($A^*_{\rm N}$)
of  SSGA (NSGA) at the epoch of the minimum of NSGA (SSGA) 
cycle (which are also given in Table~2) {\it versus} time ($i.e.$ epoch
 of minimum). 
The {\it filled circle-continuous curves}
 represent  the variations in the minimum 
of  NSGA-cycle [$A_{\rm m}$] 
and $A^*_{\rm N}$,
$i.e.$ the  value of  NSGA  at the epoch of the minimum of SSGA-cycle.
 The {\it open-circle-dashed curves} represent the variations
in the  $A_{\rm m}$ of  SSGA-cycle and $A^*_{\rm S}$,
$i.e.$ the  value of  SSGA  at the epoch of the minimum of NSGA-cycle.
The quantities at the epochs of minima of NSGA- and SSGA-cycles
 are shown by  {\it black} and {\it red} colors, respectively.
{\bf Lower panel:} The {\it black filled-circle-continuous curve}  and
the {\it red open-circle-dashed curve}
  represent the variations of North--South asymmetry in the corresponding 
aforementioned quantities 
  at the epochs of the minima of NSGA- and SSGA-cycles, respectively. 
The positive
 and negative values imply the North dominance and the South dominance,
 respectively.}
\label{f8}
\end{figure}

\begin{figure}
\centering
\includegraphics[width=12cm]{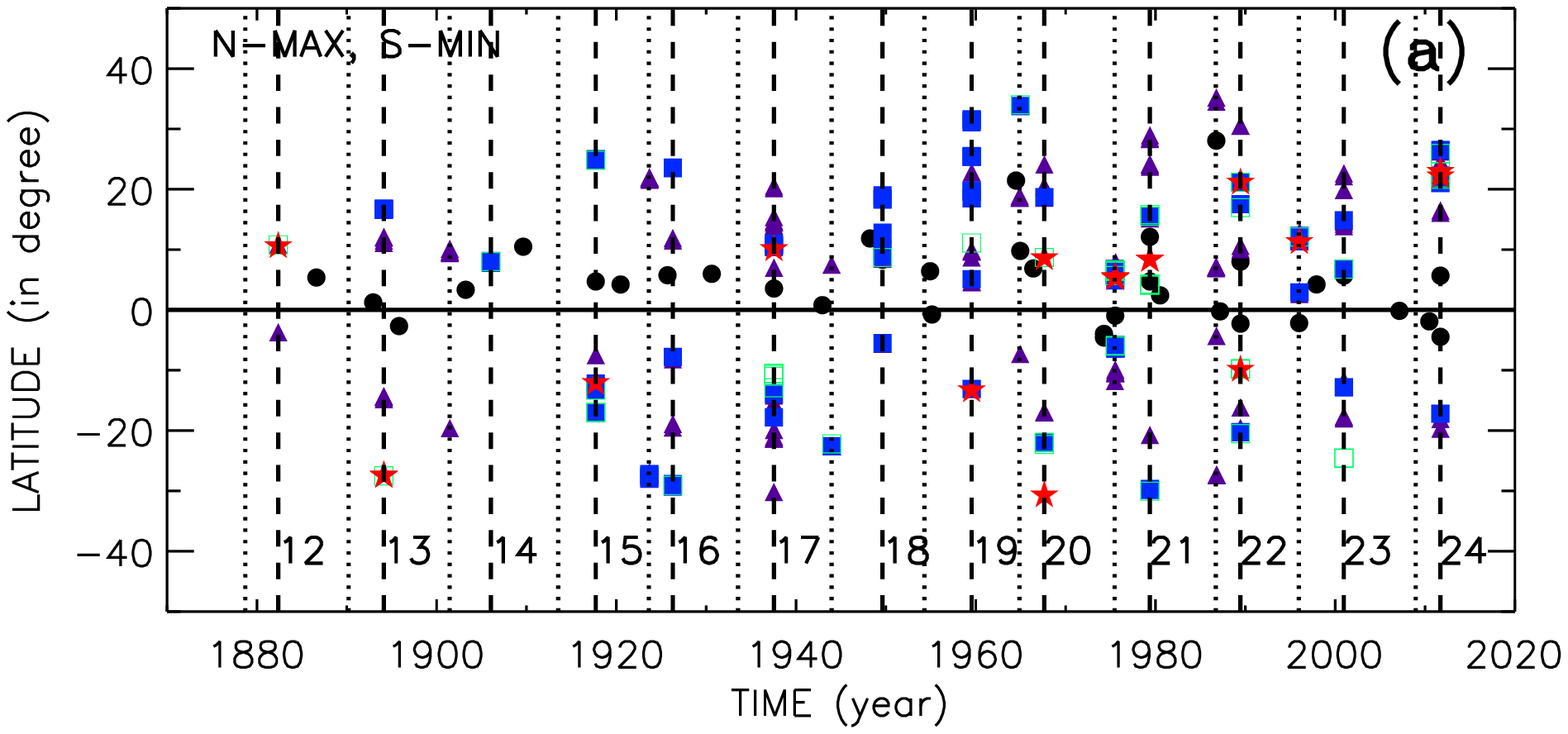}
\includegraphics[width=12cm]{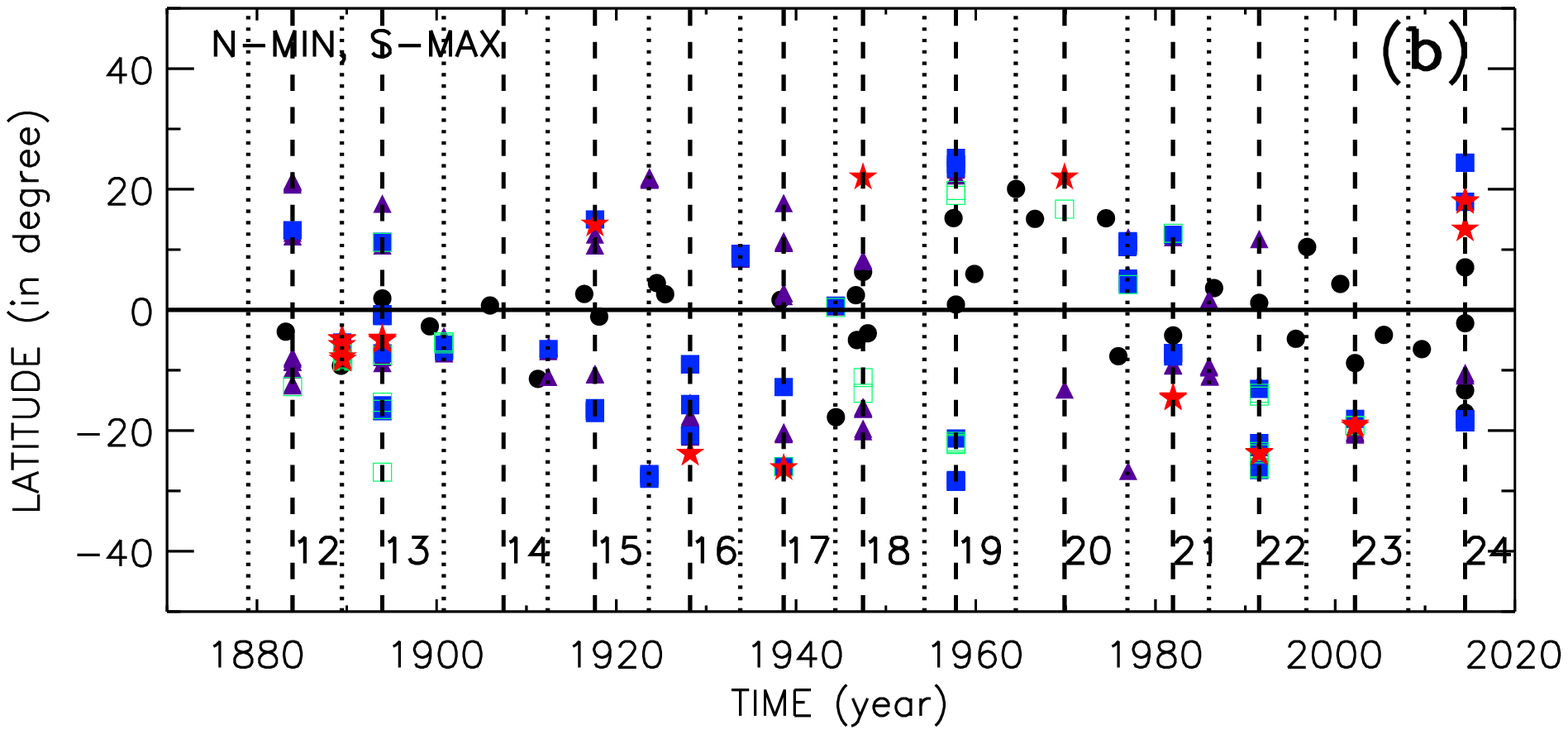}
\caption{Latitude--time distributions of sunspot groups: 
 ({\bf a}) the latitude 
distributions of sunspot groups at the epochs of  $A_{\rm M}$ of NSGA-cycles
 and at the epochs of  $A_{\rm m}$ of SSGA-cycles, and ({\bf b}) the latitude 
distributions of sunspot groups   at the epochs of  
  $A_{\rm M}$ of SSGA-cycles and at the epochs of $A_{\rm m}$ of NSGA-cycles.
 The sunspot groups having  areas 0\,--\,100 msh, 
100\,--\,200 msh, 200\,--\,300 msh, 300\,--\,400 msh, and 400\,--\,500 msh 
 are indicated by  the {\it filled-circle} (black), 
{\it filled-triangle} (light-blue), {\it filled-square} (blue), {\it open-squre}
(green), and {\it filled-star} (red), respectively. The {\it dotted-} and {\it dashed-vertical lines}
 are drawn at the epochs of the  minimum [$A_{\rm m}$] and  maximum 
 [$A_{\rm M}$], respectively. Waldmeier cycle numbers are  shown near 
the epochs of maxima of the cycles. (Note: here  the time intervals 
$ME$ to  $ME + 0.01$  and $me$ to $me + 0.1$ are used,
where $ME$ and $me$ are the epochs of maximum and minimum, respectively.)}  
\label{f9}
\end{figure}

As can be seen in the upper panel of Figure~7,
except for Cycles 12 and 13, for all of the  other cycles  
$A_{\rm M}$ of NSGA-cycle is larger than   $A^*_{\rm S}$ 
(see black curves in the upper panel, also see Table~2).
  Obviously, the mean value of the 
 asymmetry between these quantities 
suggests that the activity is  dominant in the northern hemisphere 
(see the black curve in the lower panel). The pattern of the  variation in the 
asymmetry  suggests that 
there exists an $\approx$55-year periodicity (there are large positive 
values at Cycles 14, 19, and 24).  
The pattern  of a 130\,--\,140-year cycle  that is seen in 
Figure~4(a) is not clear here. 
 Except at the Cycles 14, 15, and 20, for  the remaining  
 cycles  $A_{\rm M}$ of SSGA-cycle is considerably larger than
  $A^*_{\rm N}$  (see the red curves in the upper panel). 
 (For Cycle 15 the $A_{\rm M}$ of the SSGA-cycle is slightly smaller than
 $A^*_{\rm N}$.) Obviously, 
 the mean value of the corresponding asymmetry suggests  that the activity 
is dominant in the southern hemisphere (see the red curve in lower panel). 
There are patterns of an $\approx$55-year 
cycle and there is also a trend of a 130\,--\,140-year 
 periodicity (the values of the asymmetry has a large negative values at 
Cycles 12 and 24.)

As can be seen in the upper panel of Figure~8, the patterns of variations in 
 all of the quantities that are  shown in it are similar and 
 there are patterns  of 44\,--\,66 years periodicity in each quantity. 
Except for Cycle~19, for the remaining  cycles $A_{\rm m}$ of SSGA-cycle is 
smaller  than $A^*_{\rm N}$ (see the red curves in the upper panel, also
 see Table~2). For Cycles 12 and 16 and particularly 
for Cycles 20 and 21  the  $A_{\rm m}$ of NSGA-cycle is larger  than
the corresponding  $A^*_{\rm S}$ (see the black curves in the upper panel).
 For all of the remaining  cycles 
 $A_{\rm m}$ of the NSGA-cycle is 
smaller than  $A^*_{\rm S}$. There exists a 44\,--\,66 year periodicity in 
 both the asymmetry between the $A_{\rm m}$ of NSGA-cycle and the corresponding
 $A^*_{\rm S}$ and that between the $A_{\rm m}$ of SSGA-cycle  and
 the corresponding  $A^*_{\rm N}$ (see the lower panel of Figure~8)

Overall, the above results suggest that  a  44\,--\,66-year
 periodicity is present  in the North--South asymmetry of activity at 
the epochs of both   minima  and maxima  of NSGA- and SSGA-cycles. 
A long (130\,--\,140 years) periodicity may exist  in the 
North--South asymmetry at the 
maxima of  SSGA-cycles,  but as already mentioned in Section~3.1  
the 143 years of   data used here are inadequate to determine  the precise 
value  of this periodicity.

During the Maunder Minimum (1645\,--\,1715) sunspot activity was not 
completely absent. That is, during the late Maunder Minimum   activity
 was present in the low latitudes (between $0^\circ$ and $-20^\circ$) of  
 the southern hemisphere (see Figure~1(a) of \opencite{soko94}).  
Figure~9(a) shows the  latitude distributions (tentative) of  
sunspot groups at the epochs  of maxima of NSGA-cycles  and at the epochs of 
minima of SSGA-cycles. At these epochs the activity is 
more in the northern hemisphere (also see  Figures 7 and 8). 
Figure~9(b) is  same as Figure~9(a) but for the 
latitude distributions of sunspot groups at the epochs of maxima of 
SSGA-cycles and at the epochs of minima of NSGA-cycles.
 At  these epochs, except in a few cycles (Cycles~19, and 20),  
the activity is  greater in the southern  hemisphere
(also see Figures~7 and 8). 
(Note: The epochs of  maxima and the minima  were obtained from  the 
13-month smoothed data. In  Figure~9 we have  used the  daily data 
very close to these epochs of maxima and minima.)
The pattern of activity (in Cycles 12\,--\,18 and  Cycles 21\,--\,23) shown 
in Figure~9(b) is qualitatively similar to the pattern of activity 
 during the late Maunder Minimum (1670\,--\,1715, in about four cycles). 
 Therefore, it seems that  
during the late Maunder Minimum
activity was absent around the epochs of the maxima of  NSGA-cycles
and the minima of  SSGA-cycles, and some activity was present at the 
epochs of the maxima of some SSGA-cycles and the minima of some NSGA-cycles.
We think that during the late Maunder Minimum,  because of  unique alignments 
of the giant  planets (\opencite{jj05})  there could be  
 a large combined affect of the Sun's 
rotation and the inclination of the Sun's Equator to the Ecliptic 
 causing  equatorial crossing  of a large amount of the magnetic flux of the 
southern hemisphere.  This canceled the flux of  northern hemisphere.    
The cancellation of flux 
 might have taken place in the deeper layers of the Sun's convection zone.

\section {Conclusions and Discussion}
The North--South asymmetry in solar activity may  have
 some important implications for the solar dynamo
 mechanism. We analyzed the combined GPR and DPD daily sunspot group data
  during the period 1874\,--\,2017 and
  studied  the  North--South asymmetry in the maxima
and minima  of  Solar Cycles 12\,--\,24.
We derived the time-series of the 13-month smoothed monthly
mean corrected whole-spot areas of the sunspot groups in the Sun's whole sphere
(WSGA), northern hemisphere (NSGA), and southern hemisphere (SSGA). From these
smoothed time series we obtained the values of the maxima and minima, and the
corresponding epochs, of the WSGA, NSGA, and SSGA Cycles 12\,--\,24.
 We find that   a 44\,--\,66-year periodicity exists in the North--South
 asymmetry of minimum. A long periodicity (130\,--\,140 years) may exit  in the 
asymmetry of maximum. A statistically significant correlation exists between
  the maximum of  SSGA Cycle~$n$
 and the rise time of  WSGA Cycle~$n+2$.
A   reasonably significant
correlation also exists  between
 the maximum of  WSGA Cycle~$n$
 and the decline time of  WSGA Cycle~$n+2$.
Using these relations we obtained
  the values $11.70\pm 0.15$ years, $11.2\pm 0.2$ years,
and $11.45\pm 0.3$ years for the lengths of WSGA Cycles 24, 25, and 26,
 respectively, 
 and hence,  $2020.57 \pm 0.15$  (July 2020), $2031.77 \pm 0.2$
 (October 2031),
 and $2043.22 \pm 0.3$  (March 2043) for
the minimum epochs (start dates) of  WSGA
Cycles 25, 26, and 27, respectively.
We  also obtained $2025.41 \pm 0.51$  (May 2025) and
 $2036.18 \pm 0.51$  (March 2036)
for the maximum epochs of  WSGA Cycles 25 and 26, respectively.
Our analysis also suggests that the
  lengths and the starting times  of  ISSN Cycles 24, 25, and 26
 would be almost the same as the corresponding  WSGA-cycles.
 It seems during the late Maunder Minimum sunspot 
activity was absent around the epochs of the maxima of  NSGA-cycles
and the minima of  SSGA-cycles, and some activity was present at the 
epochs of the maxima of some SSGA-cycles and the minima of some NSGA-cycles.

The ratio of the number of large sunspot groups to the number of 
 small  sunspot groups is
generally  smaller in the minimum  than  in the maximum of a solar
 cycle. That is, 
the epochs of the  minima and maxima of solar cycles comprise relatively 
small and large numbers of large sunspot groups, respectively. 
The magnetic structures of large and small sunspot groups are rooted at
 relatively  deep and shallow layers of the solar convection 
 zone~(\opencite{jg97b}; \opencite{kmh02}; \opencite{siva03}). Hence, 
the long-term periodicities  of  the    
 North--South asymmetry in  the maxima and minima   of  solar
 cycles might originate  at relatively deep and shallow layers of the 
solar convection zone, respectively. Therefore, there could be
 differences in the long-term variations of solar  maximum and minimum   
 and in their North--South asymmetry.
  However, the physical reason for the existence of the 130\,--\,140-year  and 
44\,--\,66-year  periodicities
in the North--South asymmetry of solar activity is not clear.
 In fact, these 
periodicities seem to be a common features of many solar and solar-related 
phenomena, but the physical processes responsible for these also 
 are  not clear yet~($e.g.$ \opencite{tan11}; \opencite{gao16};
 \opencite{komit16}, and references therein). On the other hand  
there could be a connection  between  solar variability 
(both long- and short-term)
and planetary configurations~(\opencite{juc03}; \opencite{jj05}, 
\opencite{abreu12}; \opencite{cc12}; \opencite{wil13}; \opencite{chow16};
 \opencite{stef16}, and references therein). 
 
In  our earlier  studies the amount of activity 
 in the northern hemisphere around minimum and mainly   
 in the southern hemisphere  during the declining phase  of the Solar
 Cycle~$n$ correlated with the amplitude of   
 Cycle~$n+1$~(\opencite{jj07}; \citeyear{jj08}; \citeyear{jj15}).
 Here, we find that the amplitude of the
Cycle~$n$ correlate with the rise time, and anticorrelate with the 
 declining-time,  of  Cycle $n+2$. Altogether these results  suggest
that solar dynamo carries memory over at least  three solar cycles. It  is   
consistent with the  kind of flux transport dynamo model in which 
 a long magnetic memory is an important criteria~(\opencite{dg06}). 
 A low amplitude predicted for   
 Cycle~25 in \inlinecite{jj15} 
and the maximum epoch of  WSGA Cycle~25 predicted here qualitatively match 
 the recent 
predictions of the amplitude and maximum epoch of this solar cycle by 
an advanced flux transport model  based on the polar fields at cycle 
minimum (\opencite{lh18}).
However, the predictions based on the polar fields at cycle minimum
could be uncertain because it was unclear exactly when these polar-field 
measurements should be taken (\opencite{sval05}) and Cycle 24 has not 
yet ended.   The aforementioned relationships of the amplitude of WSGA-Cycle~$n$
 with rise and decline times of Solar Cycle  $n+2$ imply that 
an odd (even) cycle have a connection with its previous odd (even) cycle. 
As  in  Javaraiah (\citeyear{jj07}, \citeyear{jj08}, \citeyear{jj15}),
 here also 
we would like to suggest that  the 
 aforementioned relationships may be also related to the changes in 
the polarities of the Sun's magnetic fields and both meridional and down 
flows have roles in the magnetic-flux transport. 
Polar fields change sign about one year after the  
maximum of a cycle takes place. The polarities of the fields at the maximum
 epoch of Cycle~$n$, the declining phase of Cycle $n+1$, and the rising
 phase of Cycle~$n+2$ differ from those of the fields at the declining
 phase of Cycle~$n$, 
the rising phase of Cycle $n+1$, and the declining phase of Cycle~$n+2$. 
Therefore,  a physical reason behind   the Equation~1 may be that  a large
southern-hemisphere  
flux transported from the maximum of a large Cycle~$n$ may cancel a relatively  
 large amount of northern-hemisphere flux in  the rising phase  
(including minimum epoch) of Cycle~$n+2$ ($i.e.$ it decreases the rate of 
flux emergence) causing an increase in the rise 
time of Cycle~$n+2$.
A physical reason behind   Equation~2 may be that a large
flux transported from the maximum of a large Cycle~$n$ may cancel a relatively
 large amount of
 flux in the declining phase of Cycle~$n+2$ ($i.e.$ it enhances the rate of 
 flux decay) causing a decrease in decline time of  Cycle~$n+2$.

As per the aforementioned physical interpretation of Equation~1  even- and 
odd-numbered solar cycles  could be arranged separately according to their 
strengths as follows:
 $12 > 14 < 16 < 18 > 20 < 22 > 24 < 26 ?$ and
 $13 > 15 < 17 < 19 > 21 > 23 < 25 ?$ 
(if  Cycle~23 is considered as a reasonably small odd-numbered cycle). 
However, we think that the relationships (between two consecutive sunspot 
cycles)  that yielded a low amplitude for Cycle~25 in an our earlier 
analysis (\opencite{jj15}) are much stronger than the 
relationships (between alternate cycles) found here. In addition, at least 
at the time of a steep decrease
 in the orbital angular momentum of the Sun may have an effect on the Sun's 
rotation and hence, the corresponding change in the Sun's spin-rate 
 may be responsible for a grand minimum  type of
low activity (\opencite{jj05}). The next unique alignments of the giant 
planets and hence, the steep decrease in the Sun's orbital 
angular momentum will be taking place in 2030 (see \opencite{jj05}).     
 Cycles~25 and 26  will be taking place  in the vicinity of 2030 and 
both might be weaker than Cycle~24, as predicted by \inlinecite{jj17}.
On the other hand,  the exact physical reason behind the aforementioned all
 relations  needs to be established.

\section{Acknowledgments} 
The author thanks the anonymous referee for 
useful comments and suggestions. The author acknowledges the work of all the
 people contribute  and maintain the GPR and DPD  Sunspot databases. 
The sunspot-number data are provided by WDC-SILSO, Royal Observatory of 
Belgium, Brussels.

\section{Disclosure of Potential Conflicts of Interest} 
The author declares that he has no conflicts of interest.

{}
\end{article}
\end{document}